\documentclass[12pt,BCOR=0mm,DIV=35]{article}
\usepackage{latexsym}
\usepackage{graphicx}
\usepackage{amsmath}
\usepackage{amsfonts}
\usepackage{amssymb}

\newcommand{\Be}[1]{\begin{equation}\label{e:#1}}
\newcommand{\Ee}{\end{equation}}

\title{On the geometry of conformal mechanics}
\author{K. Andrzejewski\thanks{E-mail: k-andrzejewski@uni.lodz.pl}  , 
 J. Gonera\thanks{E-mail: jgonera@uni.lodz.pl}\\
Department of Theoretical Physics and Computer Sciences, \\
University of {\L}\'od\'z,
Pomorska 149/153, 90-236 {\L}\'od\'z, Poland}

\date{}

\begin{document}
\maketitle
\begin{abstract}
A geometric picture of conformally invariant mechanics is presented. Although the standard form of the model is recovered, 
the careful analysis of global geometry of phase space leads to the conclusion that, in the attractive case, the singularity related to
the phenomenon of "falling on the center" is spurious. This opens new possibilities concerning both the interpretation  and quantization of the model.
Moreover, similar modification seem to be relevant in supersymmetric and multidimensional generalization of conformal mechanics.
\end{abstract}
Conformal mechanics \cite{b:1} and its supersymmetric extensions \cite{b:2}  provide simple yet nontrivial examples of
conformally invariant theories. There appeared numerous  papers dealing with various aspects of these models \cite{b:3}.
An interesting geometrical interpretation of the conformal mechanics in terms of nonlinear sigma  model on the $SO(2,1)$  
group has been given in the nice
paper by Ivanov et. al. \cite{b:4} (see \cite{b:5} for supersymmetric counterpart of such approach). The  authors used the method of
 nonlinear realizations  \cite{b:6}  together  with exponential parameterization of $SO(2,1)$  group to define a covariant  dynamics
 on group manifold (actually -- on appropriate  coset space). The dynamics results from setting all Cartan forms, except one, equal
to zero. This leads, through the so-called inverse Higgs phenomenon \cite{b:7}, to the equation  of motion which, by a
 simple change of variables, is shown to be equivalent to standard equation of conformal mechanics.
The methods developed by Ivanov, Krivonos and Leviant may be applied in other contexts. For example, it has been used in 
Ref. \cite{b:8} to construct Galilean conformally  invariant dynamics in arbitrary space-time dimension.
\par
One of the important ingredients of the approach proposed in Ref. \cite{b:4} is the use of exponential parameterization of the $SO(2,1)$ group.
However, it is known that, in the case of semisimple group, the exponential parameterization does not provide 
the map covering the whole group. For this reason some global topological properties on conformal dynamics can escape our attention.
\par
In the present paper we propose an alternative approach to conformal dynamics based on the Kirillov method of orbits \cite{b:9} which allow us to classify the sympletic manifolds on which a given group acts transitively preserving the sympletic structure.  When  applied to the  $SO(2,1)$ group, 
Kirillov method yields quite simple construction of general conformal dynamics  grasping all details of its global structure.
It appears that the model is well-defined for all values of coupling constant including the negative ones and provides an interesting starting point for generalizations and quantization. 
\par
We start with $SO(2,1)$ group consisting of $3\times3$ matrices ${\Lambda ^\mu}_\nu$ obeying
$g_{\mu\nu}{\Lambda^\mu}_\nu{\Lambda^\nu}_\beta=g_{\alpha\beta}$; we adopt the convention:
$g_{\mu\nu}=diag(+,-,-)$. Its Lie algebra reads 
\Be{1}
[M_\mu,M_\nu]=-i{\varepsilon_{\mu\nu}}^\alpha M_\alpha,
\Ee
with $\epsilon_{012}=1$. Upon identifying 
\Be{2}
K=M_0+M_1,\quad H=M_0-M_1,\quad D=M_2,
\Ee
the conformal algebra is recovered
\Be{3}
[D,H]=-iH,\quad [D,K]=iK,\quad [K,H]=-2iD.
\Ee

According to the general theory \cite{b:9} in order to classify all $SO(2,1)$--invariant sympletic manifolds 
 on which $SO(2,1)$ acts transitively it is sufficient to find the orbits of coadjoint action 
(in fact, for semisimple groups it is equivalent to classify the orbits of adjoint action).
In our case the coadjoint action reads 
\Be{4}
\zeta_\alpha'={(\Lambda^{-1})^\beta}_\alpha\zeta_\beta,
\Ee
where $\zeta_\alpha$ are the coordinates in linear space dual to $SO(2,1)$ algebra. The invariant  
(degenerate) Poisson structure is defined by
\Be{5}
\{\zeta_\alpha,\zeta_\beta\}=-{\varepsilon_{\alpha\beta}}^\gamma\zeta_\gamma.
\Ee
The generators of conformal algebra are represented by the following functions (cf. eq. (\ref{e:2}))
\Be{6}
K=\zeta_0+\zeta_1,\quad H=\zeta_0-\zeta_1,\quad  D=\zeta_2.
\Ee
The Hamiltonian equations of motion 
\Be{7}
\dot\zeta_\alpha=\{\zeta_\alpha,H\},
\Ee
take the form 
\Be{8}
\dot\zeta_0=-\zeta_2,\quad \dot\zeta_1=-\zeta_2,\quad \dot \zeta_2=\zeta_1-\zeta_0,
\Ee
and yield 
\begin{align}
\label{e:9}
&\zeta_0(t)=\frac{1}{2}(\zeta_0-\zeta_1)t^2-\zeta_2t+\zeta_0,\nonumber\\
&\zeta_1(t)=\frac{1}{2}(\zeta_0-\zeta_1)t^2+\zeta_2t+\zeta_1,\\
&\zeta_2=(\zeta_1-\zeta_0)t+\zeta_2,\nonumber
\end{align}
where $\zeta_\alpha=\zeta_\alpha(0)$.
To classify the invariant sympletic structures one has to find all orbits. This is an easy task. There are three families of orbits:
\begin{enumerate}
\item[(i)] upper ($\zeta_0>0$) and lower ($\zeta_0<0$) sheets of two-sheeted hyperboloids
\Be{10}
\zeta^\alpha\zeta_\alpha=\lambda^2>0,
\Ee
\item[(ii)] one-sheeted hyperboloids
\Be{11}
\zeta^\alpha\zeta_\alpha=-\lambda^2<0,
\Ee
\item[(iii)] forward ($\zeta_0>0$) and backward ($\zeta_0<0$) cones
\Be{12}
\zeta^\alpha\zeta_\alpha=0,
\Ee
\end{enumerate}
and the trivial orbit, $\zeta_\alpha=0$. The Poisson structure (\ref{e:5}), when restricted to these orbits, becomes nondegenerate.
\par
Let us analyze successively the relevant structures.  We start with the orbits (\ref{e:10}) and consider first the  upper sheet. 
Introduce new variables, $0<x<\infty$, $-\infty<p<\infty$, by 
\begin{align}
\label{e:13}
&\zeta_0=\frac{p^2}{4m}+\frac{\lambda^2}{mx^2}+\frac{mx^2}{4},\nonumber\\
&\zeta_1=\frac{-p^2}{4m}-\frac{\lambda^2}{mx^2}+\frac{mx^2}{4},\\
&\zeta_2=-\frac{1}{2}xp;\nonumber
\end{align}
here $m$ is an arbitrary positive constant (we are dealing with dimensionless coordinates but this can be easily cured).
One can check that eqs. (\ref{e:13}) provide a smooth map for our manifold. Moreover, $x$ and $p$ are global Darboux coordinates, $\{x,p\}=1$. 
Hamiltonian takes the form 
\Be{14}
H=\frac{p^2}{2m}+\frac{2\lambda^2}{mx^2},
\Ee
so the standard form of conformal mechanics is recovered.
\par
As far as the lower sheet, $\zeta_0<0$, is concerned we note that 
$\zeta_0\rightarrow -\zeta_0$, $\zeta_1\rightarrow -\zeta_1$, $\zeta_2\rightarrow \zeta_2$
is an automorphism of Poisson structure.
Therefore, one obtains the relevant modification of the mapping (\ref{e:13}) which leads to negative definite  Hamiltonian (due to $\zeta_0<0$ and $|\zeta_0|>|\zeta_1|$):
\Be{15}
H=-\frac{p^2}{2m}-\frac{2\lambda^2}{mx^2}.
\Ee
The case $(ii)$ is much more interesting. The one-sheeted hyperboloid provide the manifold which cannot be covered by one map. We choose the following maps
\Be{e:16}
\begin{array}{l}
U_1=\{(\phi,\zeta_0)|\quad  0<\phi<2\pi,\quad -\infty<\zeta_0<\infty\},\\
U_2=\{(\phi',\zeta_0')|\quad  -\pi<\phi'<\pi,\quad -\infty<\zeta_0'<\infty\};
\end{array}
\Ee
here $\phi$ $(\phi')$ denotes the polar angle on $\zeta_1,\zeta_2$ plane.
Their intersection consists of two open sets  $U_1\cap U_2=V_1\cup V_2$
\Be{e:17}
\begin{array}{l}
V_1=\{(\phi,\zeta_0)|\quad  0<\phi<\pi,\quad -\infty<\zeta_0<\infty\},\\
V_2=\{(\phi',\zeta_0')|\quad  \pi<\phi'<2\pi,\quad -\infty<\zeta_0'<\infty\},
\end{array}
\Ee
with transition functions
\Be{e:18}
\begin{array}{l}
\zeta_0'=\zeta_0\quad \rm{ on }\quad V_1\cup V_2, \\
\phi'=\left\{
\begin{array}{lll}
\phi&\rm{on}& V_1 ,\\
\phi-2\pi& \rm{on}& V_2.
\end{array} \right.
\end{array}
\Ee
Note that both maps provide (local) Darboux coordinates. 
In order to make the contact with standard conformal mechanics consider the intersection of our hyperboloid with the  plane
$\zeta_0+\zeta_1=0$.  It consists of two straight lines   $\zeta_0+\zeta_1=0$, $\zeta_2=\pm \lambda$. Consider two manifolds:
 $M_{\pm}=\{\zeta^\alpha\zeta_\alpha=-\lambda^2|\ \zeta_0+\zeta_1
\genfrac{}{}{0pt}{2}{>}{<}0\}$. Together with two lines defined above they cover the whole hyperboloid (cf. Fig. 1)
\begin{center}
\begin{tabular}{c}
\includegraphics{obrazki-4.ps}+\\
\verb+Figure 1+\\
\end{tabular}
\end{center}
In order to parameterize $M_+$  consider the mapping obtained from eq. (\ref{e:13}) by replacing $\lambda^2\rightarrow -\lambda^2$
\begin{align}
\label{e:19}
&\zeta_0=\frac{p^2}{4m}-\frac{\lambda^2}{mx^2}+\frac{mx^2}{4},\nonumber\\
&\zeta_1=\frac{-p^2}{4m}+\frac{\lambda^2}{mx^2}+\frac{mx^2}{4},\\
&\zeta_2=-\frac{1}{2}xp.\nonumber
\end{align}
One can check that eq. (\ref{e:19}) provides a diffeomorphism between the half-plane $x>0$,  $-\infty<p<\infty$  and $M_+$;  Moreover, $x$ and $p$ are Darboux
coordinates. Again, using the Poisson algebra automorphism $\zeta_0\rightarrow -\zeta_0$, $\zeta_1\rightarrow -\zeta_1$, $\zeta_2\rightarrow \zeta_2$ one 
finds the  diffeomorphism between the left half-plane $x<0$, $-\infty<p<\infty$ and $M_-$
\begin{align}
\label{e:20}
&\zeta_0=-\frac{p^2}{4m}+\frac{\lambda^2}{mx^2}-\frac{mx^2}{4},\nonumber\\
&\zeta_1=\frac{p^2}{4m}-\frac{\lambda^2}{mx^2}-\frac{mx^2}{4},\\
&\zeta_2=-\frac{1}{2}xp;\nonumber
\end{align}
$M_+$ nad $M_-$ cover the whole hyperboloid except two straight  lines $\zeta_0+\zeta_1=0$. Let us find the motion in terms of $x,p$  variables.
 To this end we find from eq. (\ref{e:9})
\Be{21}
\zeta_0(t)+\zeta_1(t)=(\zeta_0-\zeta_1)t^2-2\zeta_2t +(\zeta_0+\zeta_1).
\Ee
Equation $\zeta_0(t)+\zeta_1(t)=0$ has two solutions for $\zeta_0-\zeta_1\neq0$ and one for $\zeta_0-\zeta_1=0$. So the solution traversers
$M_+$ or $M_-$ in finite time depending on whether the energy $E=\zeta_0-\zeta_1$  is negative or positive, respectively; the only exception is zero-energy motion.
Taking into account that the allowed region is $E\geq\frac{-2\lambda^2}{mx^2}$ for $x>0$ and $E\leq\frac{2\lambda^2}{mx^2}$ for $x<0$ we arrive at the picture
of the motion in terms of $x$ variable which is illustrated on Fig. 2.
\begin{center}
\begin{tabular}{c}
\includegraphics{obrazki-2.ps}+\\
\verb+Figure 2+\\
\end{tabular}
\end{center}
We conclude that description of dynamics in terms of positive values of coordinate $x$ is incomplete. The singularity related to the effect of "falling on the center in finite time"
is spurious. It is an artefact of the choice of coordinates in sympletic manifold. The situation is somewhat similar  to that encountered in general relativity. For example, the only
real singularity  in Schwarzschild solution resides at the center; however, with the standard choice of coordinates the metrics diverges at the horizon.
\par
Finally, consider the case of light cone. The forward ($\zeta_0>0$) and backward ($\zeta_0<0$)  cones are separated by one point which by itself forms an  orbit.
 The intersection of both cones with the plane $\zeta_0+\zeta_1$ form now one straight  line. Eq. (\ref{e:21}) can be rewritten in form
\Be{22}
\zeta_0(t)+\zeta_1(t)=(\zeta_0-\zeta_1)\left(t-\frac{\zeta_2}{\zeta_0-\zeta_1}\right)^2.
\Ee
So $\zeta_0(t)+\zeta_1(t)\geq 0$ ($\zeta_0(t)+\zeta_1(t)\leq 0)$ for forward (backward) cone. The points $\zeta_0=\zeta_1$, $\zeta_2=0$ are fixed points of dynamics.
 According to eq. (\ref{e:22}) any other trajectory crosses the line $\zeta_0+\zeta_1=0$  exactly once. This depicted on Fig. 3.
\begin{center}
\begin{tabular}{c}
\includegraphics{obrazki-3.ps}+\\
\verb+Figure 3+\\
\end{tabular}
\end{center}
Let us consider the forward cone. It can be parameterized by $\zeta_1$ and $\zeta_2$ variables running over the plane with the origin deleted.
Define the canonical variables  $x,p$ by 
\Be{23}
\begin{array}{l}
\zeta_1=-\frac{p^2}{4m}+\frac{mx^2}{4},\\
\zeta_2=-\frac{1}{2}xp,
\end{array}
\Ee
for $-\infty <x,p<\infty$.
\par To analyze this mapping we introduce the complex variables 
\Be{24}
\begin{array}{l}
\zeta=\zeta_1+i\zeta_2,\\
u=\frac{\sqrt{m}}{2}x-\frac{ip}{2\sqrt{m}}.
\end{array}
\Ee
Then eqs. (\ref{e:23}) read 
\Be{25}
\zeta=u^2.
\Ee
Consider $\zeta_1,\zeta_2$ plane with the cut along negative $\zeta_1$-axis and the origin deleted. Note that the cut corresponds to the line $\zeta_0+\zeta_1=0$.
Typical trajectory (with $\zeta_0,\zeta_1$) is drawn on Fig. 4. 
\begin{center}
\begin{tabular}{c}
\includegraphics{obrazki-1.ps}+\\
\verb+Figure 4+\\
\end{tabular}
\end{center}
Eqs. (\ref{e:19}) imply
\Be{26}
\zeta=\left(\sqrt{\frac{E}{2}}\left(t-\frac{\zeta_2}{\zeta_0-\zeta_1}\right)-i\sqrt{\frac{E}{2}}\right)^2.
\Ee
Assume we start on the upper sheet. By comparing  eqs. (\ref{e:24})--(\ref{e:26}) one finds
\Be{27}
\begin{array}{l}
x=-\sqrt{\frac{2E}{m}}\left(t-\frac{\zeta_2}{\zeta_0-\zeta_1}\right),\\
p=-\sqrt{2mE}.
\end{array}
\Ee
Similarly, starting from the lower sheet one finds solutions with the minus sign on the right-hand side replaced by plus.
\par
As a last issue let us discus briefly the conformal  symmetry of  our dynamics. The  Hamiltonian  form of the action of conformal group is described 
using standard methods of Hamiltonian dynamics. The important point is that the Hamiltonian is an element of nontrivial Lie algebra.
 Therefore, the symmetry generators depend explicitly on time. They read 
\begin{align}
\label{e:28}
&H(t)=H,\nonumber\\
&D(t)=D+tH,\\
&K(t)=K+2tD+t^2H,\nonumber
\end{align}
$H(t),D(t)$ and $K(t)$ are the integrals of motion and obey (in the sense of Poisson brackets) the conformal algebra. The relevant canonical 
symmetry transformations coincide with conformal ones. The details will be published elsewhere \cite{b:10}.
\par
{\bf Acknowledgments}\
We are obliged  to P. Kosi\'nki and P. Ma\'slanka for stimulating discussions.
We also acknowledge a support from  MNiSzW grant No. N202331139  (J.G.), as well as from the  earmarked subsidies MNiSzW for Young Scientists grant No. 545/199 (K.A.).

\end{document}